\documentclass[conference,10pt,letterpaper]{IEEEtran}
\IEEEoverridecommandlockouts

\usepackage{cite}
\usepackage{graphicx}
\usepackage{amsmath,amssymb,amsfonts}
\usepackage{textcomp}
\usepackage{xcolor}
\usepackage{booktabs}
\usepackage{siunitx}
\usepackage{balance}
\usepackage{stfloats}
\usepackage{float}
\usepackage{printlen}
\usepackage{paralist}

\usepackage[font=footnotesize,labelfont=bf,labelsep=period]{caption}
\usepackage[font=footnotesize]{subcaption}
\usepackage[hidelinks]{hyperref}

\makeatletter
\newcounter{runinbib}
\renewcommand{\theruninbib}{\arabic{runinbib}}
\AtBeginDocument{%
  \renewenvironment{thebibliography}[1]{%
    \section*{\refname}%
    \footnotesize
    \setlength{\parindent}{0pt}%
    \setlength{\parskip}{0pt}%
    \setcounter{runinbib}{0}%
    \def\@listctr{runinbib}%
    \def\@bibitem##1{%
      \stepcounter{runinbib}%
     \Hy@raisedlink{\hyper@anchorstart{cite.##1}\hyper@anchorend}%
      \if@filesw
        \immediate\write\@auxout{\string\bibcite{##1}{\theruninbib}}%
      \fi
      \ifnum\value{runinbib}=1
        \noindent
      \else
        \unskip\hskip 0.6em plus 0.1em
      \fi
      {\bfseries[\theruninbib]}\,%
      \ignorespaces}%
    \def\@lbibitem[##1]##2{%
      \stepcounter{runinbib}%
      \Hy@raisedlink{\hyper@anchorstart{cite.##2}\hyper@anchorend}%
      \if@filesw
        \immediate\write\@auxout{\string\bibcite{##2}{\theruninbib}}%
      \fi
      \ifnum\value{runinbib}=1
        \noindent
      \else
        \unskip\hskip 0.6em plus 0.1em
      \fi
      {\bfseries[\theruninbib]}\,%
      \ignorespaces}%
    \ignorespaces
  }{%
    \par
  }%
}
\makeatother

\title{A Process-Aware Hybrid Si/IGO Monolithic-3D 6T SRAM with BEOL Pass-Gates for the 2\,nm Node}
\author{%
  \IEEEauthorblockN{%
    Po-Chuan Wang$^{1\dagger}$,
    Dongwon Jang$^{1\dagger}$,
    Eknath Sarkar$^{1}$,
    Alexei Svizhenko$^{2}$,
    Md.~Nahid Haque Shazon$^{1}$,\\
    Piyush Kumar$^{1}$,
    Suman Datta$^{1}$, and
    Azad Naeemi$^{1}$}
  \IEEEauthorblockA{%
    $^{1}$School of Electrical and Computer Engineering,
    Georgia Institute of Technology, Atlanta, GA 30332, USA.\\
    $^{2}$Synopsys, Sunnyvale, CA 94085, USA.
    $^{\dagger}$Equal Contributions, 
    Email: pwang429@gatech.edu}}
\begin{document}
\maketitle
 
\setlength{\abovecaptionskip}{4pt}   % was 6.0pt  — image-to-caption gap
%----------------------------------------------------------------------
\begin{abstract}
We propose a monolithic-3D (M3D) 6T SRAM at the 2nm node, integrating BEOL IGO pass-gates (PGs) with an all-silicon nanosheet latch, buried power rails (BPRs), and Ru interconnects. TCAD calibrated to a state-of-the-art double-gate IGO transistor with a tri-layer HfO$_2$/ZrO$_2$/HfO$_2$ (HZH) gate stack is combined with virtual fabrication and 3D parasitic extraction to realize the first process-aware layout of this topology. A novel neighbor-cell shared source/drain (S/D) bitline (BL) design enlarges the IGO contact area to mitigate contact resistance and restore PG drive without area penalty. The resulting cell achieves a 25\% footprint reduction vs the high-performance (HP) 122 Si baseline while maintaining robust static noise margin (SNM) over a wide supply voltage range. At the 128$\times$256 subarray-level, it reduces write delay by 42.2\% and EDP by 9.7\% compared to the high-density (HD) 111 Si baseline, owing to reduced cell parasitics and wordline (WL) loading from the smaller footprint.
\end{abstract}

%======================================================================
\vspace{-10pt}   % pull the next row up
\section{\textbf{Introduction}}\label{sec:intro}
\begin{figure*}[p]
  \begin{minipage}[t]{0.31\textwidth}\centering
    \includegraphics[width=\linewidth]{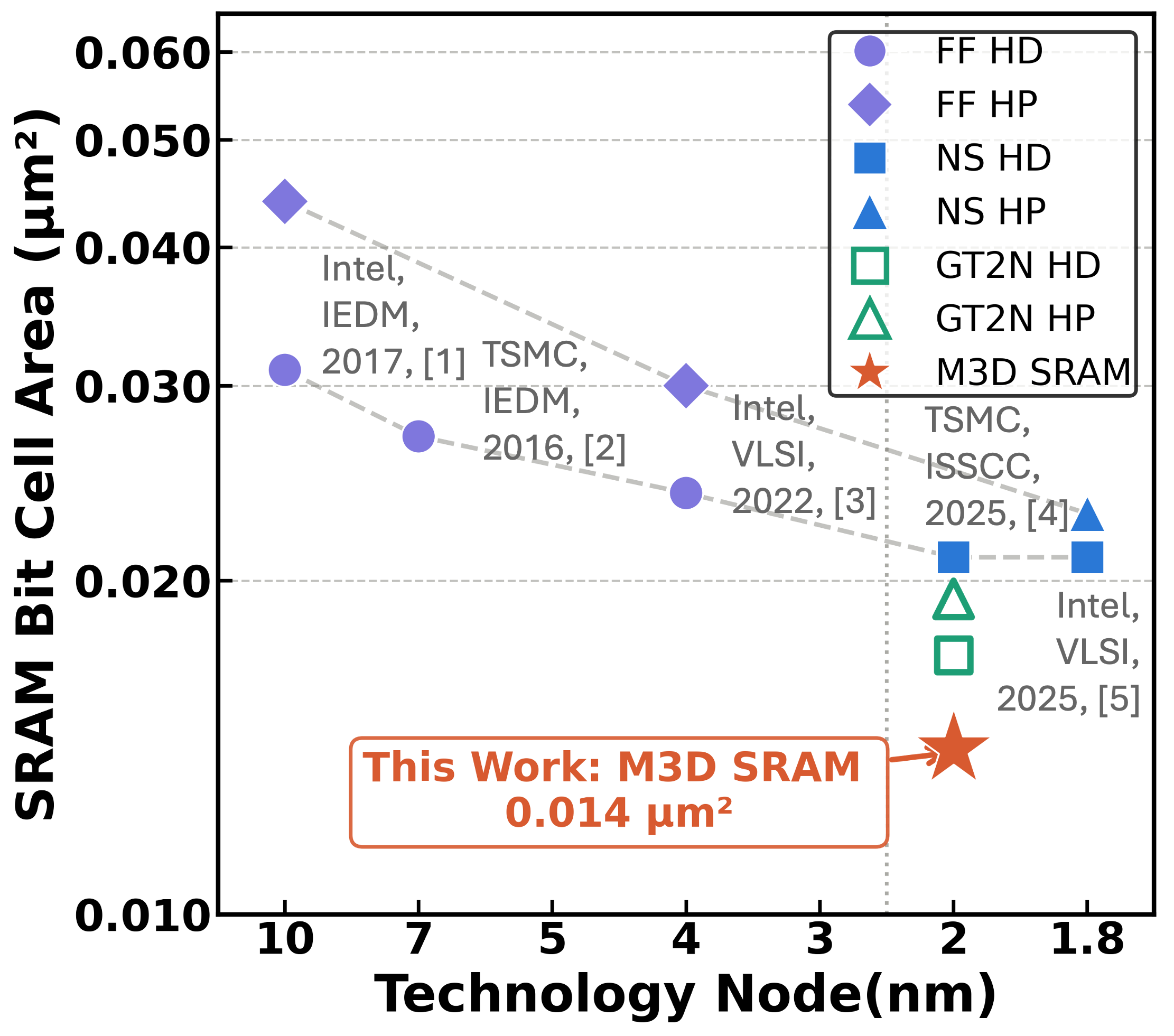}
    \caption{SRAM bitcell area vs technology node, with published HD/HP cell areas and this work's M3D and GT2N HD/HP baselines.}
    \label{fig:node_trend}
  \end{minipage}\hfill
  \begin{minipage}[t]{0.38\textwidth}\centering
    \includegraphics[width=\linewidth]{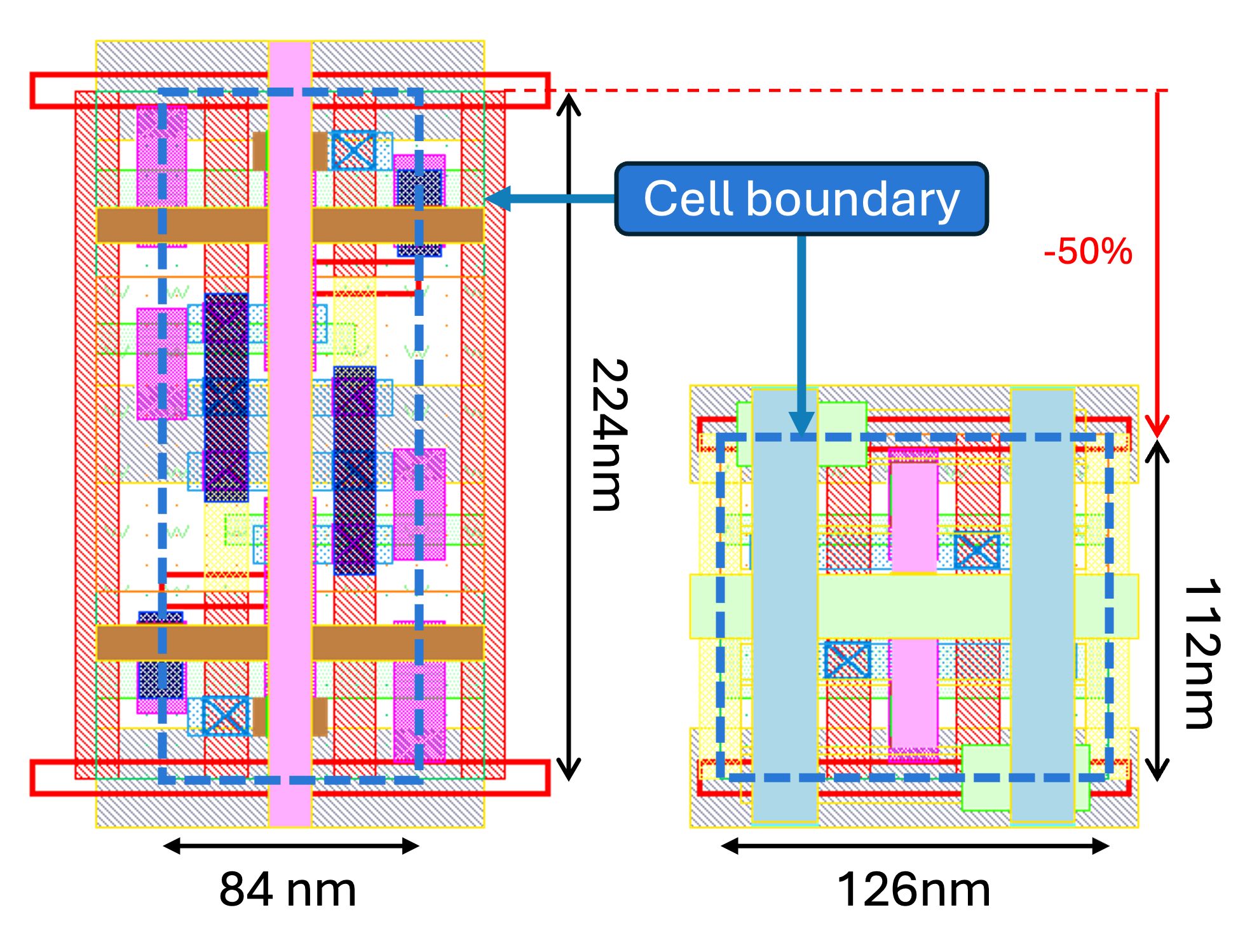}
    \caption{Bitcell area comparison of the proposed M3D cell against the 2nm Si-HP baseline, with cell width and height annotated. M3D achieves 25\% area reduction 
    vs HP.}
    \label{fig:area_red}
  \end{minipage}\hfill
    \begin{minipage}[t]{0.27\textwidth}\centering
    \includegraphics[width=\linewidth]{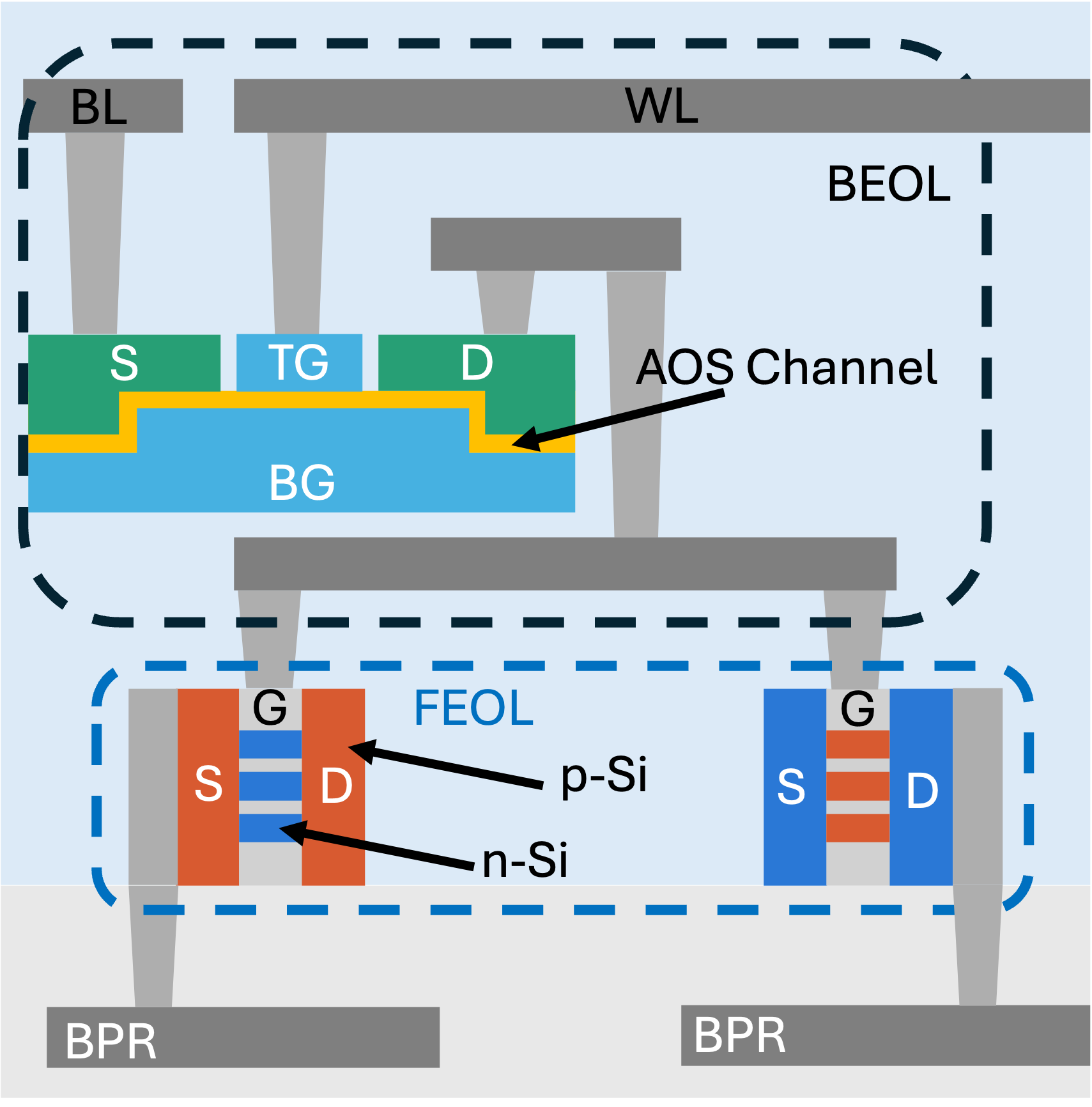}
    \caption{Proposed 6T SRAM cell: FEOL Si-nanosheet latch with two
    PGs implemented as BEOL double-gate IGO AOSTFTs.}
    \label{fig:concept}
  \end{minipage}\hfill

\end{figure*}

\vspace{-10pt}   % pull the next row up
SRAM bitcell area scaling has slowed markedly at advanced nodes, as per-node area
reduction flattens toward the 2nm generation and beyond (Fig.~\ref{fig:node_trend}).
Conventional all-Si 6T SRAMs use 111 high-density (Si-HD) and 122 high-performance
(Si-HP) sizing for different density and performance targets~\cite{auth17,wu16,bel22,chang22,fis25},
yet both are constrained by planar placement of all six transistors in the front-end-of-line
(FEOL). Monolithic-3D (M3D) integration provides an additional scaling dimension by relocating devices
vertically, enabling further footprint reduction beyond conventional planar SRAM layouts
(Fig.~\ref{fig:area_red}).

Amorphous oxide-semiconductor thin-film transistors (AOSTFTs) are
attractive for back-end-of-line (BEOL) integration owing to their low
process temperature and BEOL compatibility. However, prior M3D studies commonly 
target mature Si nodes or adopt simplified device and interconnect models, 
limiting their relevance to aggressively scaled SRAM at advanced nodes, where oxide-device characteristics, contact resistance, and BEOL parasitics become increasingly critical.

In this work, we replace the two pass-gates (PGs) in a 2nm 6T SRAM with double-gate (DG)
Ga-doped In$_2$O$_3$ (IGO) transistors using a TCAD model calibrated to experimental data,
while retaining the cross-coupled Si-nanosheet latch implemented based on the GT2N 2nm
PDK~\cite{gt2n}. Moving the PGs to the BEOL frees FEOL area for stronger
pull-down (PD) devices while maintaining a compact footprint. We further address contact-resistance through a source/drain (S/D) contact-sharing scheme between neighboring cells. 
The proposed M3D structure is constructed through virtual fabrication, followed by 3D field-solver parasitic extraction.

%======================================================================
\section{\textbf{Device Architecture and Calibration}}\label{sec:device}
The hybrid M3D architecture is shown in Fig.~\ref{fig:concept}: the
cross-coupled latch remains entirely in silicon, while both IGO access
devices are placed in the BEOL. Confining the oxide devices to the access path retains a conventional Si-nanosheet storage core and preserves the symmetry of the Si latch, which is essential for static noise margin (SNM).
At the same time, relocating the PGs to the BEOL enables independent
optimization of the oxide access devices without directly constraining
the FEOL latch geometry.

\begin{figure*}[p]

    \begin{minipage}[t]{0.215\textwidth}\centering
    \includegraphics[width=\linewidth]{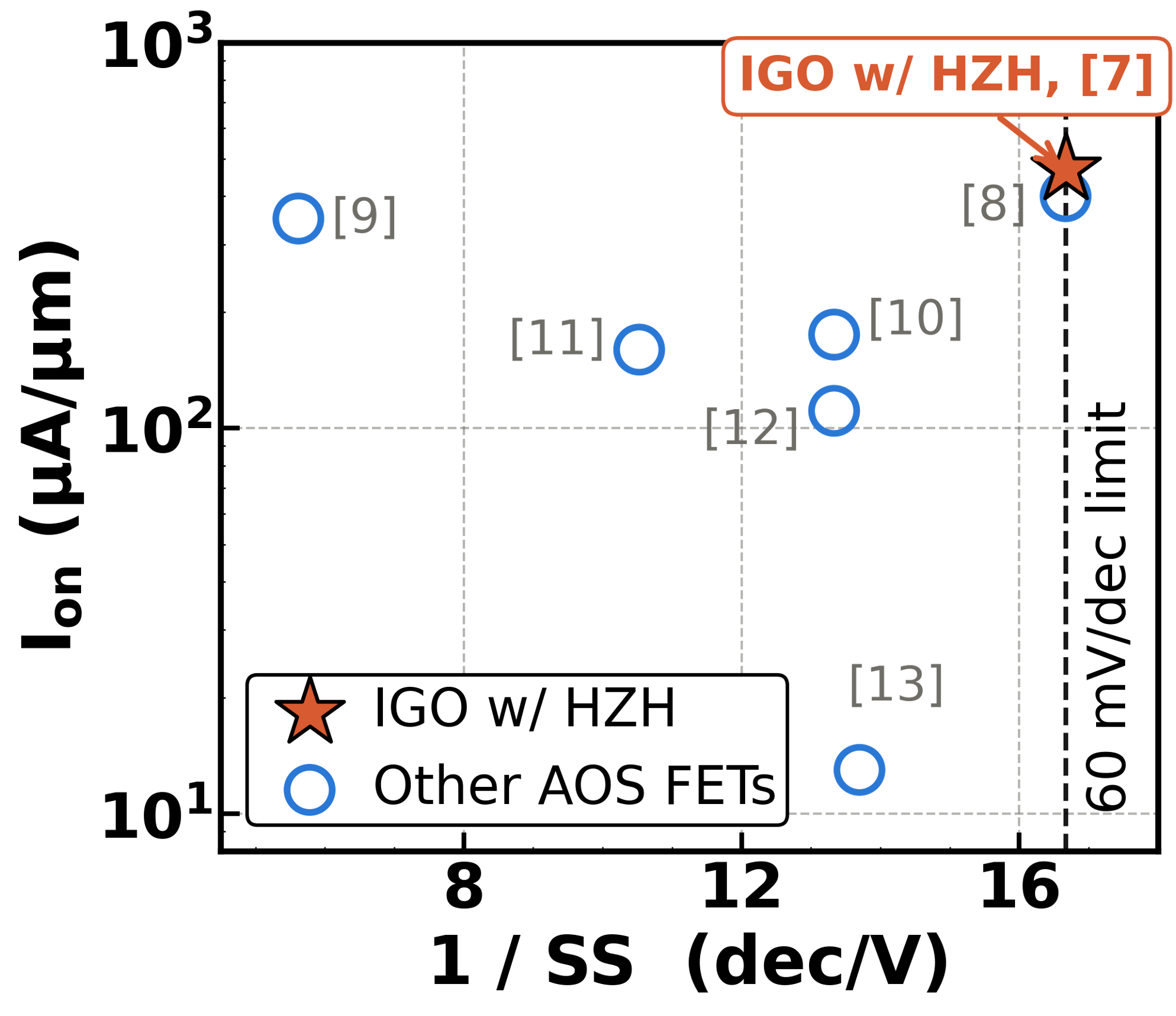}
    \caption{Survey of our double-gate IGO device~\cite{sarkar25}, against other published AOS FETs, highlighting its competitive $I_{on}$ and $SS$.}
    \label{fig:SOTA}
  \end{minipage}\hfill
      \begin{minipage}[t]{0.33\textwidth}\centering
    \includegraphics[width=\linewidth]{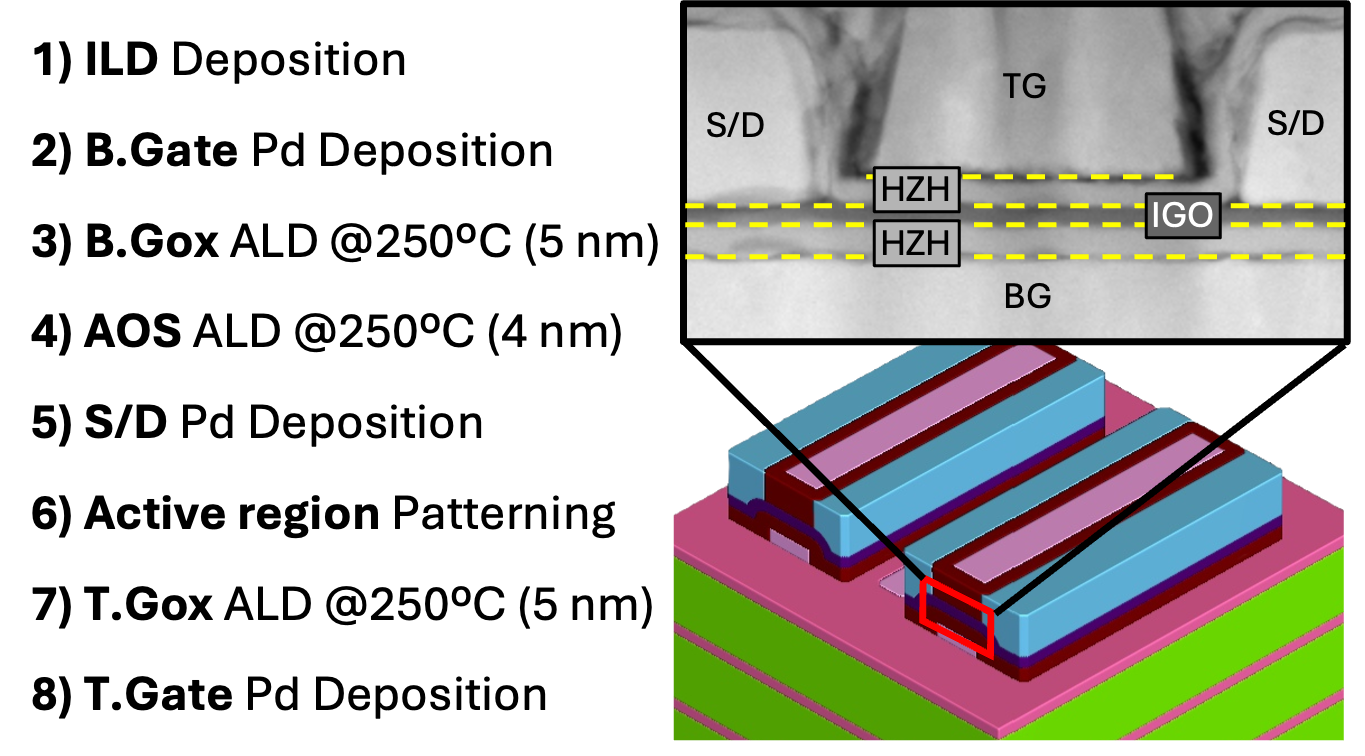}
    \caption{Fabrication process flow, TEM image, and virtually fabricated 3D structure of the experimental double-gate IGO TFT with the HZH gate stack (TG: top gate; BG: back gate; S/D: source/drain).}
    \label{fig:tem}
  \end{minipage}\hfill
  \begin{minipage}[t]{0.425\textwidth}\centering
    \includegraphics[width=\linewidth]{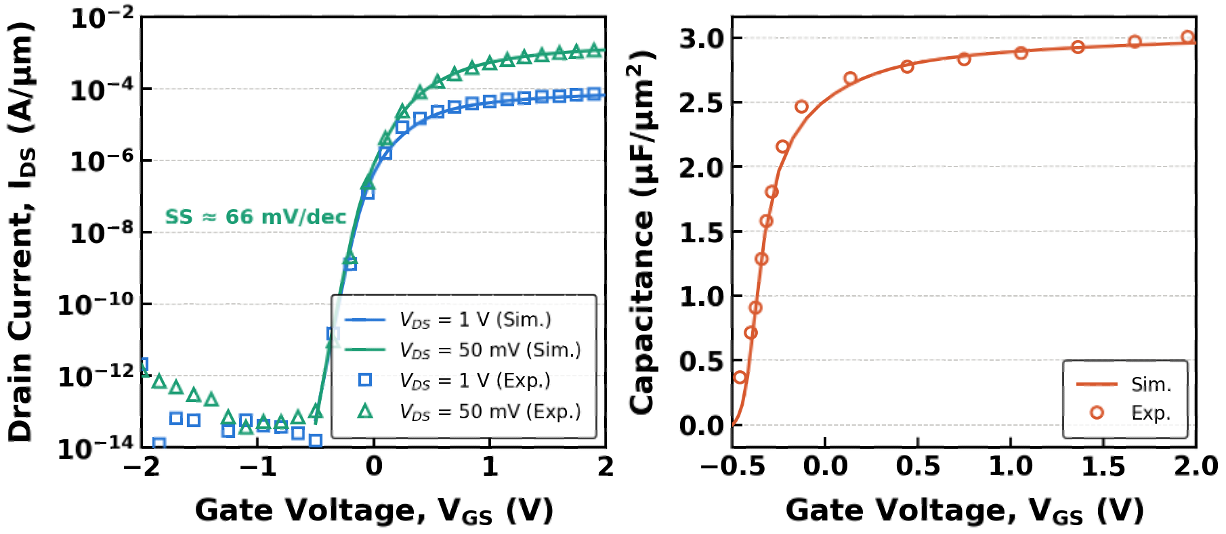}
    \caption{TCAD calibration to the measured $I_d$--$V_g$ characteristics across $V_{DS}=50$\,mV--1\,V and $C$--$V$ characteristics of
    the double-gate IGO device~\cite{sarkar25}, showing close agreement between simulation and experiment.}
    \label{fig:calib}
  \end{minipage}\hfill

\end{figure*}

We calibrate a TCAD model to a state-of-the-art DG IGO AOSTFT with a
tri-layer HZH gate stack~\cite{sarkar25} (Figs.~\ref{fig:SOTA} and~\ref{fig:tem}), which provides $V_t$ tunability, near-ideal
$SS\!\approx\!60$\,mV/dec, high drive current, and suppressed BTI.
The TCAD model is calibrated to the measured $I_d$--$V_g$ and $C$--$V$ of the DG IGO device (Fig.~\ref{fig:calib}) Starting from this calibrated model, we evaluate a 2nm-compatible PG design that retains the channel-to-gate stack, with $V_t$ tuned for sufficient drive (Fig.~\ref{fig:2nmIdVg}).
A BSIM-CMG compact model is fitted to the projected characteristics to
capture the transport and device capacitance of the oxide transistor~\cite{mben25} for subsequent
cell- and array-level simulations.

%======================================================================
\section{\textbf{Neighbor-Cell Shared-Contact Scheme}}\label{sec:share}
\begin{figure*}[p]
\vspace*{-2pt}   % ← pull this row UP, adjust value until tight
  \begin{minipage}[t]{0.39\textwidth}\centering
    \includegraphics[width=\linewidth]{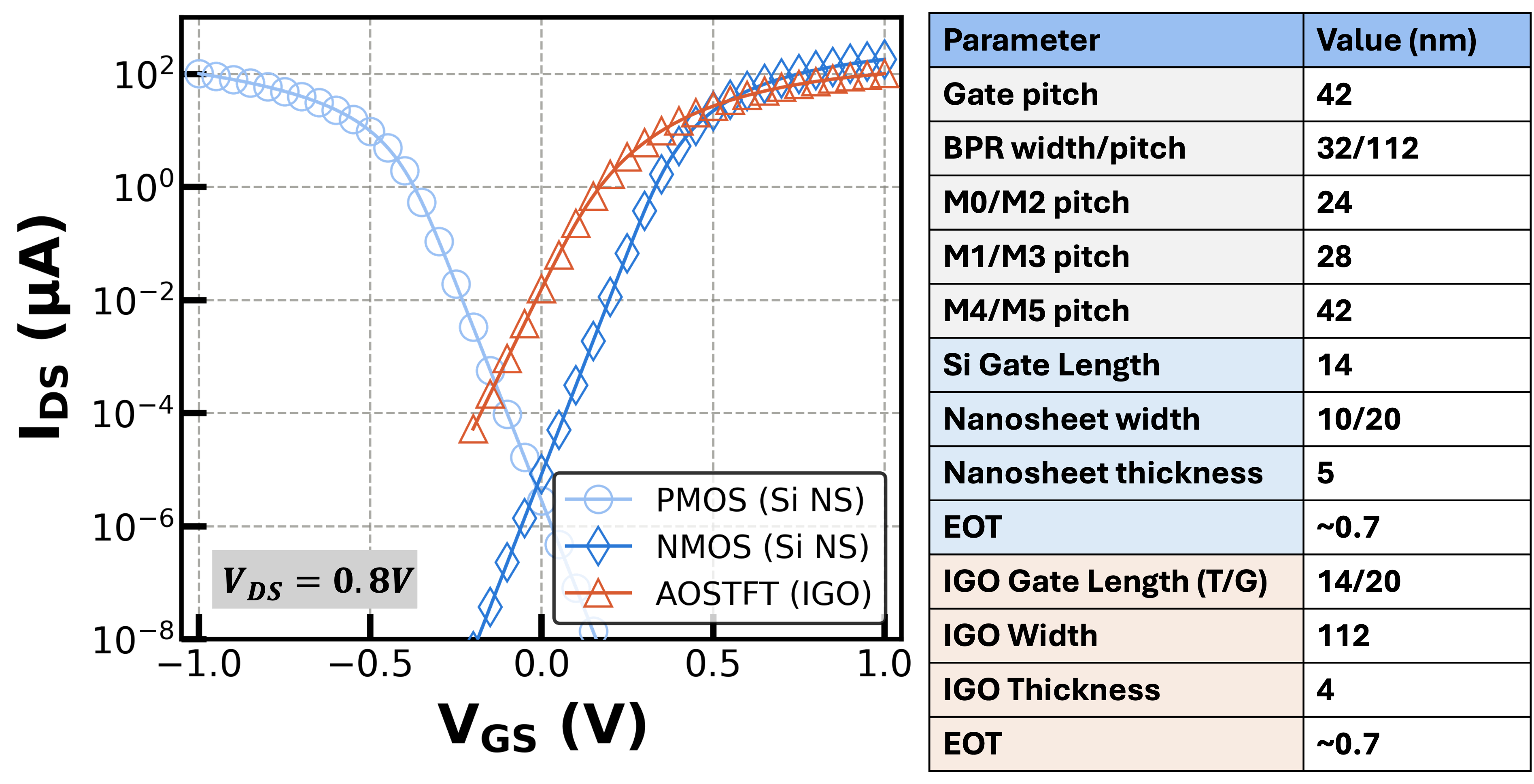}
    \caption{TCAD-evaluated 2nm-compatible IGO PG design derived from the experimentally calibrated double-gate IGO device, with table listing key 2nm design parameters. $V_t$ is tuned for sufficient PG drive for SRAM functionality~\cite{mben25}}.
    \label{fig:2nmIdVg}
  \end{minipage}\hfill
  \begin{minipage}[t]{0.22\textwidth}\centering 
    \includegraphics[width=\linewidth]{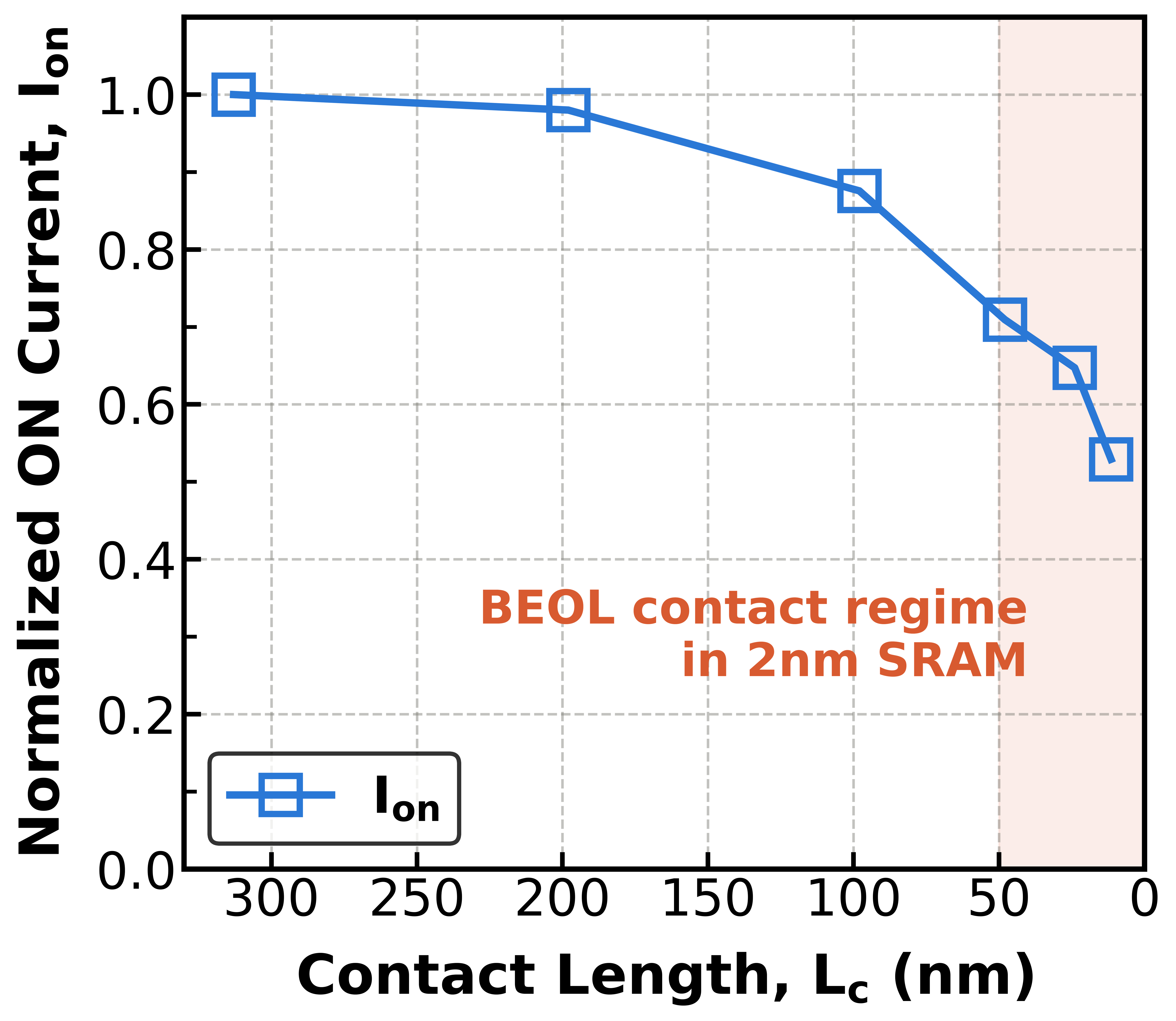}
    \caption{Normalized $I_{on}$ vs IGO S/D contact length, illustrating the drive degradation that motivates the shared-contact scheme.}
    \label{fig:Rc_scaling}
  \end{minipage}\hfill
 \begin{minipage}[t]{0.35\textwidth}\centering
    \includegraphics[width=\linewidth]{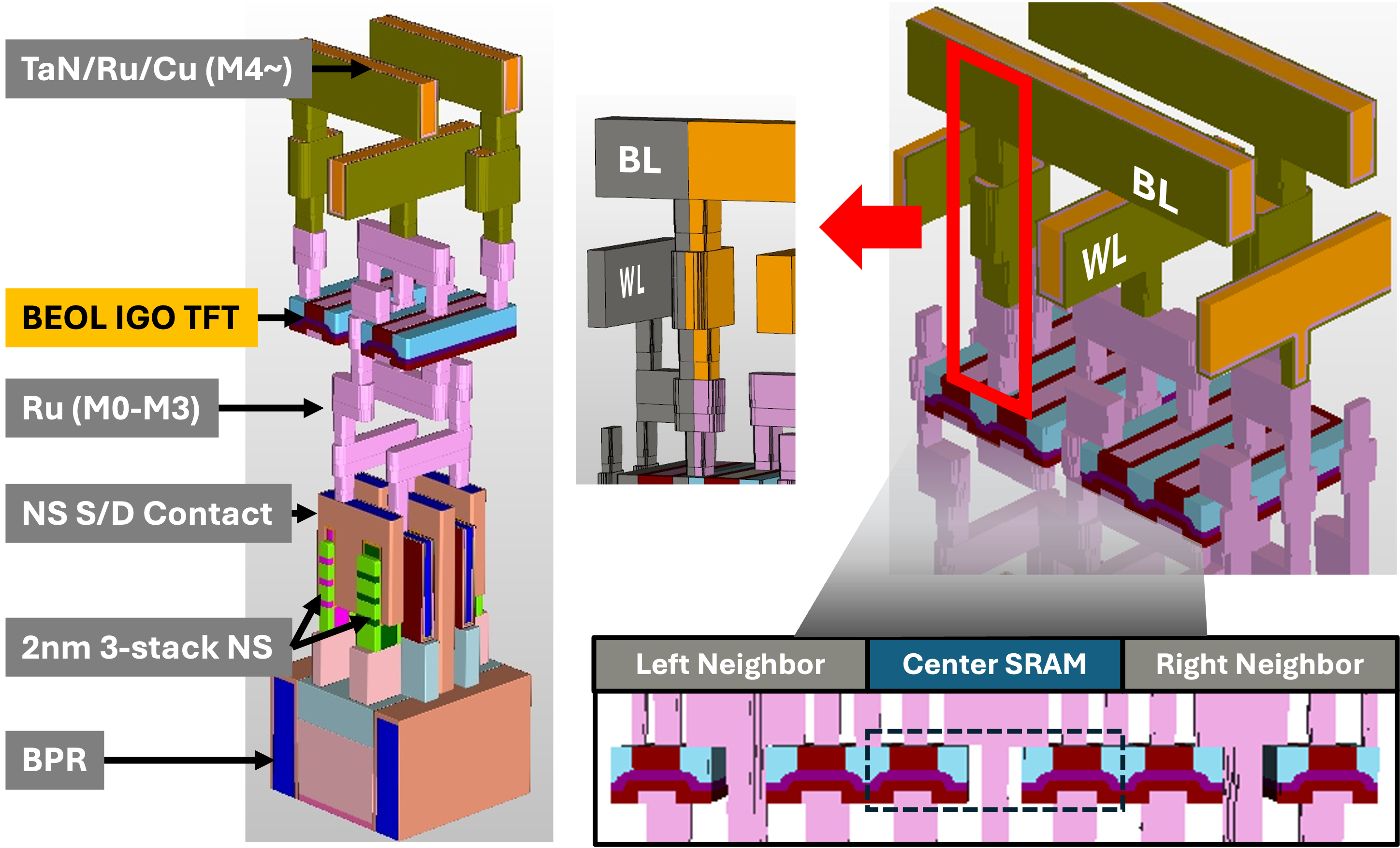}
    \caption{Virtually fabricated M3D 6T SRAM structure and the shared-contact scheme. The center bitcell shares its BL-S/D contacts with neighboring cells, enlarging effective contact area and reducing contact resistance.
    }
    \label{fig:share}
  \end{minipage}\hfill
\end{figure*}

The IGO S/D contact resistance is strongly influenced by the
metal--semiconductor Schottky barrier and is partially reduced by the
extended back-gate. However, in a compact SRAM cell at advanced nodes,
the available BEOL contact length can be as small as 10-20\,nm, below
the transfer length, causing the contact resistance to increase sharply
and degrade PG drive (Fig.~\ref{fig:Rc_scaling}). We enlarge the effective
contact area by sharing the BL-side S/D contact between neighboring
cells on a common BL node (Fig.~\ref{fig:share}). This scheme improves
the tradeoff between high $I_\mathrm{on}$ and compact layout area.

%======================================================================
\section{\textbf{Cell Layout and 3D Virtual Fabrication}}\label{sec:area}
\begin{figure*}[p]
  \begin{minipage}[t]{0.41\textwidth}\centering
    \includegraphics[width=\linewidth]{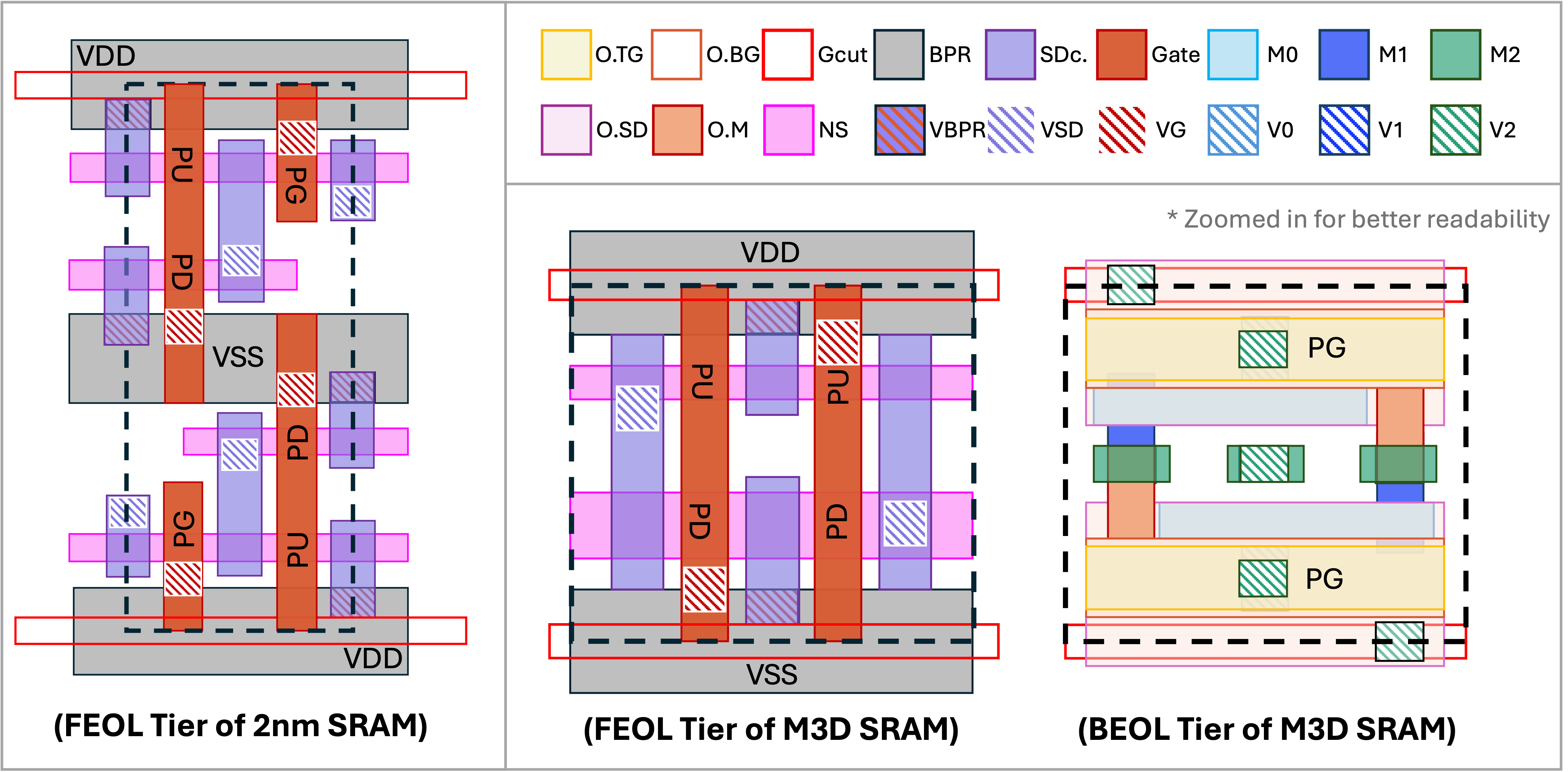}
    \caption{Cell layouts under 2nm design rules for the all-Si baselines and the proposed IGO M3D cell. 
    The M3D cell places both PGs in the BEOL tier. (O.TG/O.BG: IGO top/bottom gate; O.SD: IGO S/D; O.M: IGO interconnect metal.)}
    \label{fig:layout}
  \end{minipage}\hfill
  \begin{minipage}[t]{0.37\textwidth}\centering
    \includegraphics[width=\linewidth]{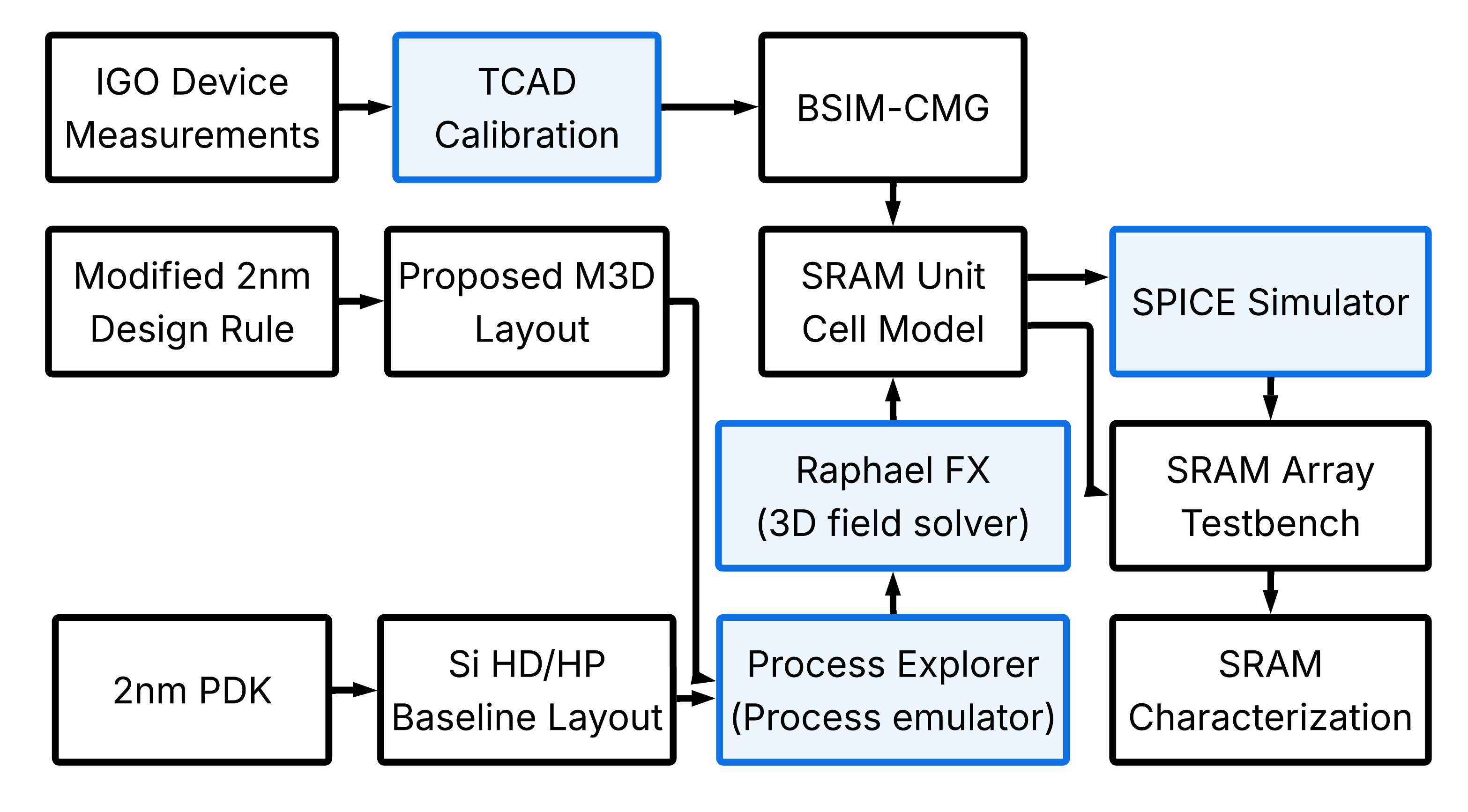}
    \caption{Simulation framework. Measured IGO data are calibrated in TCAD and captured in BSIM-CMG. Layouts are virtually fabricated in Process Explorer, with 3D parasitics extracted 
    using Raphael FX, and evaluated in SPICE.}
    \label{fig:workflow}
  \end{minipage}\hfill
\begin{minipage}[t]{0.2\textwidth}\centering
    \includegraphics[width=\linewidth]{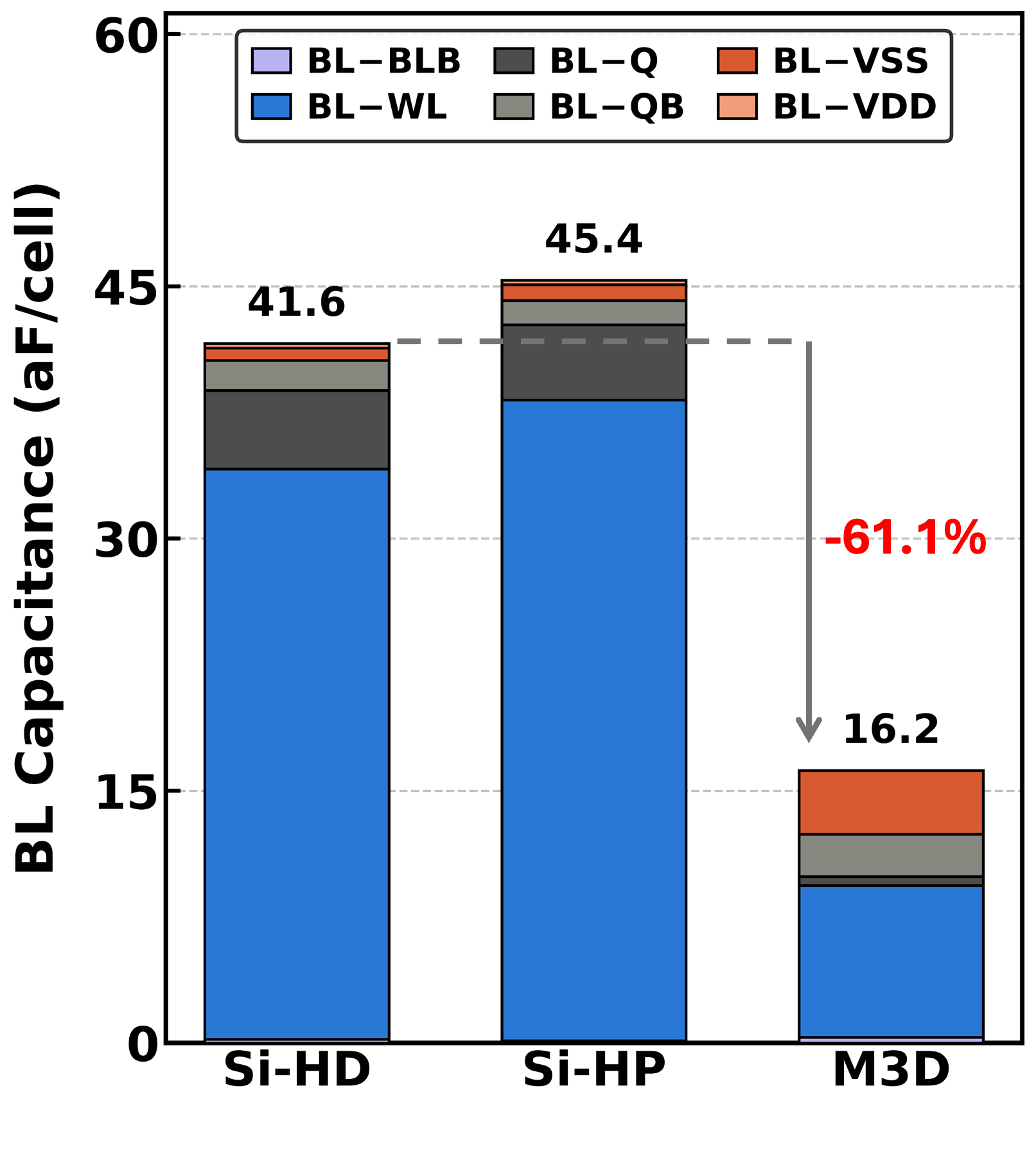}
    \caption{Extracted BL capacitance and component breakdown for all cells, showing reduced coupling in the M3D cell. }
    \label{fig:CBL}
  \end{minipage}\hfill
\end{figure*}

Fig.~\ref{fig:layout} shows the cell layouts under 2nm design rules for
the Si baselines and the proposed IGO M3D 6T cell. By relocating the PGs
to the BEOL, the silicon area beneath them is freed. Even with 20\,nm-wide
PDs, the M3D cell remains 25\% smaller than the Si-HP baseline and 18\%
smaller than the Si-HD baseline, as shown in Fig.~\ref{fig:area_red}.
Because the IGO PG and Si PD have different $I_d$--$V_g$
characteristics, $\beta$ is not a fixed constant as in an all-Si cell,
but the selected 20\,nm PD maintains a favorable PD-to-PG strength across the operating region. Crucially, the read/write balance can be tuned
through the IGO $V_t$ and independently adjustable BEOL device width.

\begin{figure*}[p]
\vspace{-10pt}
   \begin{minipage}[t]{0.5\textwidth}\centering
    \includegraphics[width=\linewidth]{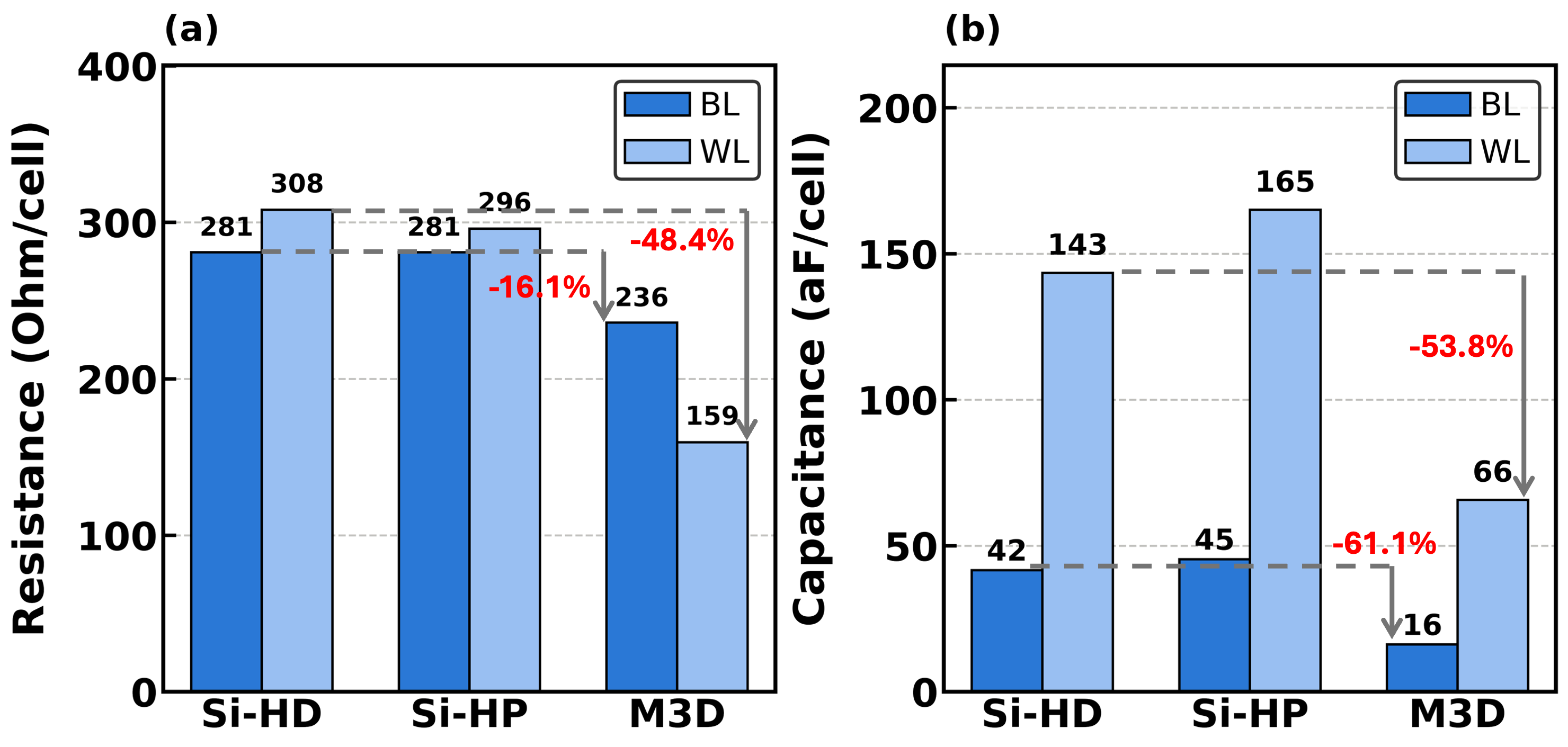}
    \caption{Extracted parasitics: (a) total resistance and (b) total capacitance of BL and WL nets for Si-HD, Si-HP, and M3D. The M3D cell reduces resistance and capacitance through shorter routing and reduced capacitive coupling.}
    \label{fig:ex_par}
  \end{minipage}\hfill
  \begin{minipage}[t]{0.48\textwidth}\centering
    \includegraphics[width=\linewidth]{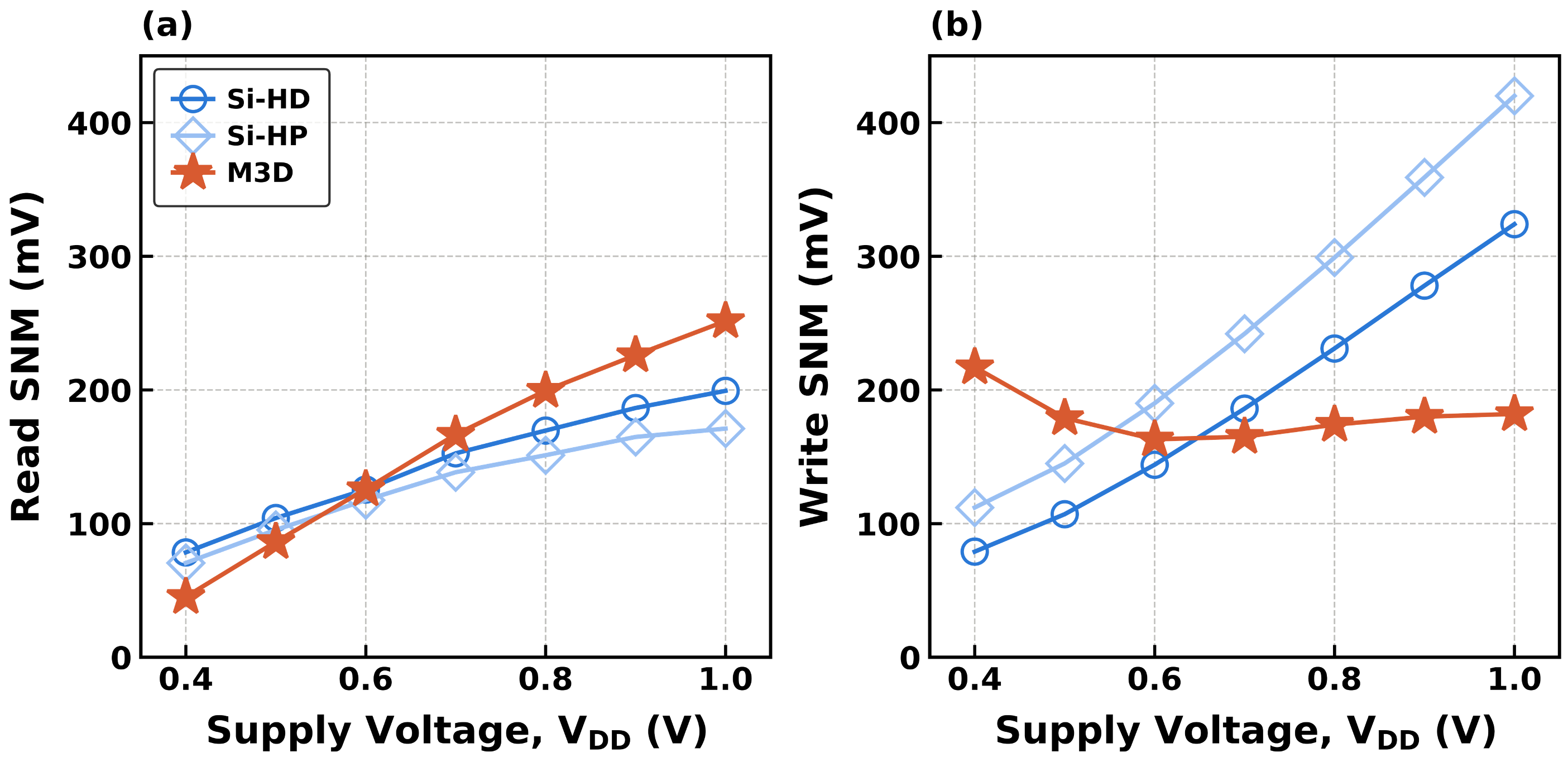}
    \caption{Static noise margin (SNM) vs supply voltage  $V_{DD}$. (a) RSNM and (b) WSNM for Si-HD, Si-HP, and M3D. 
    The M3D cell remains writable across the full $V_{DD}$ range and shows the highest WSNM at low $V_{DD}$.}
    \label{fig:snm}
  \end{minipage}\hfill
\end{figure*}

The overall workflow is shown in Fig.~\ref{fig:workflow}. The complete
M3D structure is constructed through process-aware virtual fabrication
(Fig.~\ref{fig:share})~\cite{gt2n} before parasitic extraction. A 3D
field-solver is essential to accurately capture inter-tier and inter-cell
coupling directly from the 3D structure. Figs.~\ref{fig:CBL}
and~\ref{fig:ex_par} show the extracted parasitics for the three cell
types. The proposed M3D implementation exhibits markedly lower net parasitics because its smaller footprint shortens local interconnects, while vertical separation of the BEOL PGs from the FEOL devices and redistribution of routing along the $z$-direction reduce capacitive coupling and lateral routing congestion.

\section{\textbf{SRAM Metrics: Stability, Speed, Energy}}\label{sec:metrics}

\subsection{Read and Write Stability}
In the low-voltage regime ($V_{DD}\leq0.6$\,V), the M3D cell
provides the highest WSNM of the three cells, as shown in Fig.~\ref{fig:snm}.
This advantage arises because the IGO PG write current is relatively
insensitive to $V_{DD}$, whereas the Si pull-up (PU) strengthens with
$V_{DD}$, yielding an almost flat WSNM. Above 0.6\,V, the Si
baselines exhibit higher WSNM, but all three cells retain sufficient
write margin ($>$150\,mV). The M3D cell's RSNM is lower than the Si
baselines at low $V_{DD}$ but exceeds both at high $V_{DD}$.
Consequently, a balanced operating window exists around
$V_{DD}=0.5$--$0.7$\,V, where the M3D cell maintains robust read and
write margins.

\subsection{Speed and Energy}
\begin{figure*}[p]
  \vspace{-7pt}   % pull the next row up
  \begin{minipage}[t]{0.3\textwidth}\centering
    \includegraphics[width=\linewidth]{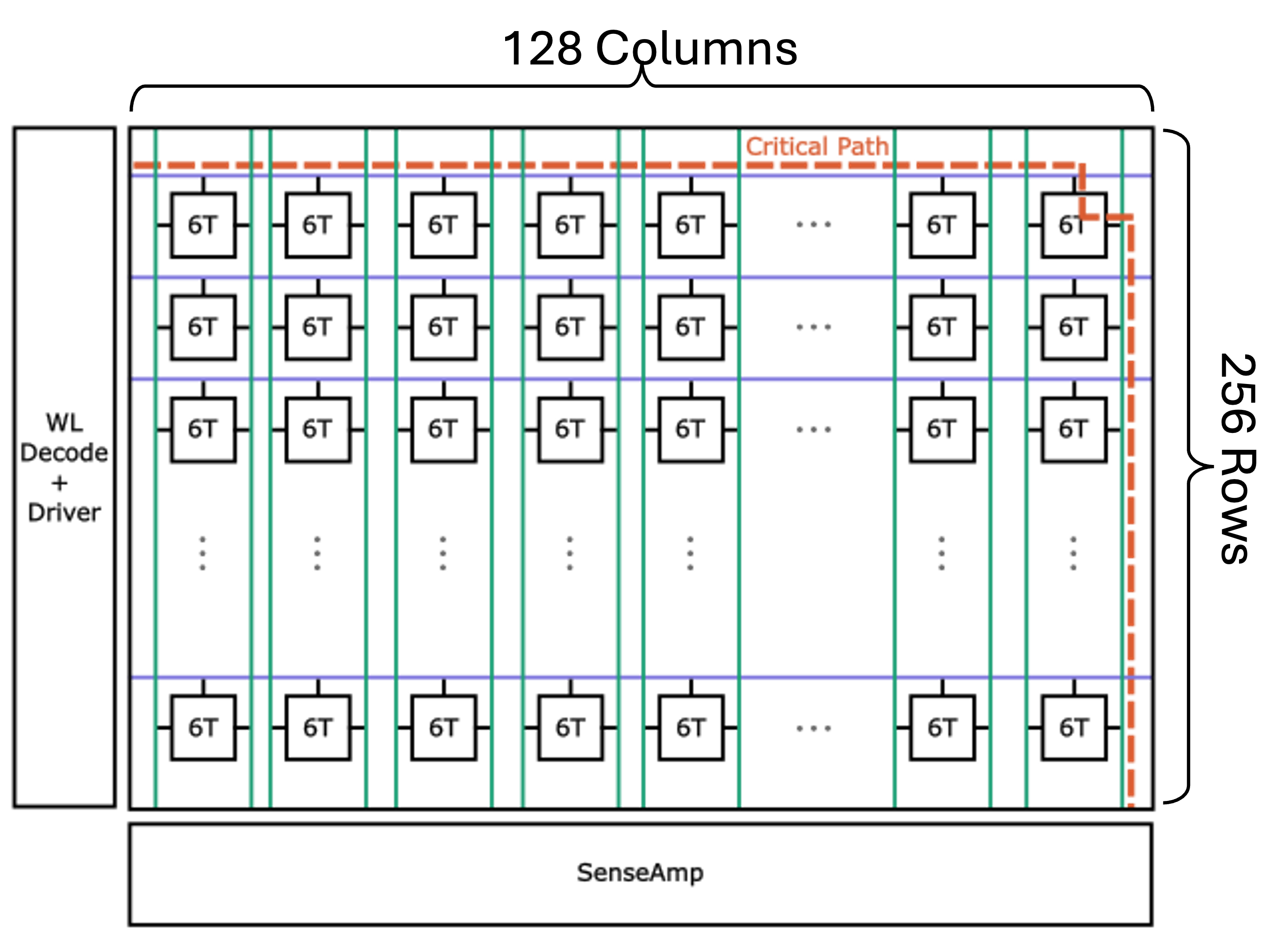}
    \caption{128-column $\times$ 256-row SRAM sub-array used for array-level timing and energy evaluation.}
    \label{fig:subarray}
  \end{minipage}\hfill
  \begin{minipage}[t]{0.69\textwidth}\centering
    \includegraphics[width=\linewidth]{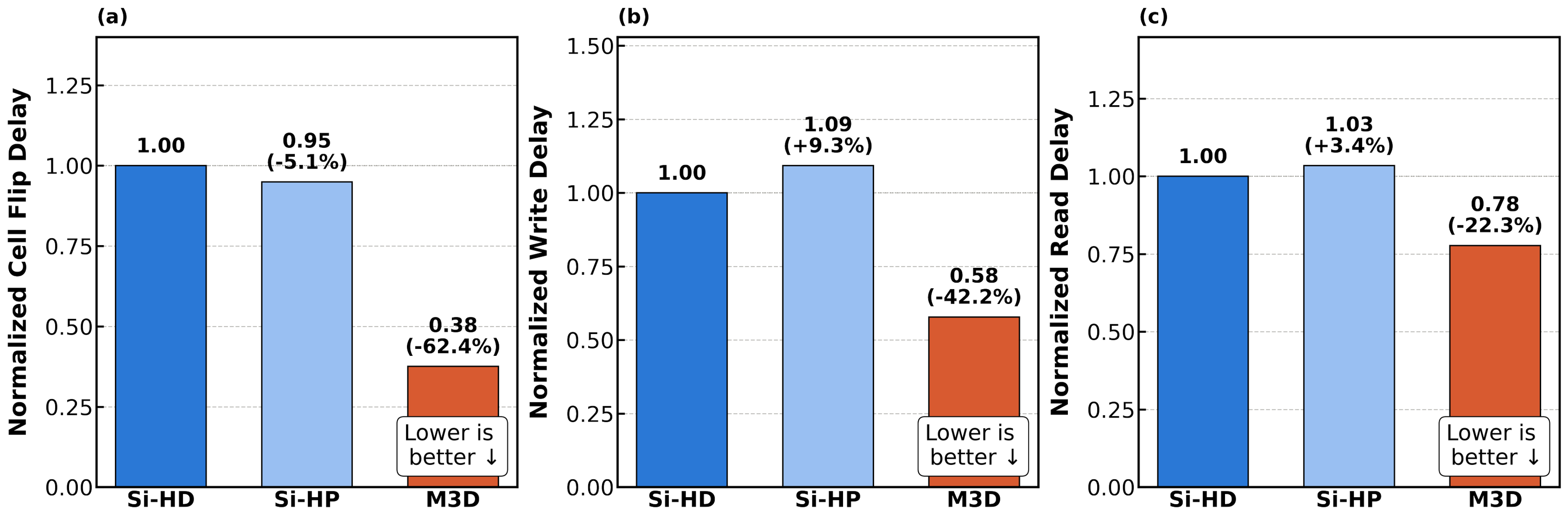}
    \caption{(a) Cell-flip delay, (b) write delay, and (c) read delay, normalized to Si-HD.
    Cell-flip delay quantifies the intrinsic storage-node transition time and excludes the WL/BL RC contributions included 
    in the array-level write delay. The M3D cell reduces write delay by $\sim$42\% and read delay by $\sim$22\% relative to Si-HD.}
    \label{fig:rw_time}
  \end{minipage}\hfill
\end{figure*}

Read/write performance is evaluated on the critical path of a
128$\times$256 subarray (Fig.~\ref{fig:subarray}), with results normalized
to the Si-HD baseline (Fig.~\ref{fig:rw_time}). Owing to its compact
footprint and layout optimization, the M3D cell reduces write delay by
$>$41\% and read delay by $>$22\% vs both Si baselines.
Notably, Si-HP shows a 5.1\% lower intrinsic cell-flip delay than
Si-HD, yet a 9.3\% higher array-level write delay. Its larger footprint
increases accumulated WL/BL interconnect length and capacitive load,
which outweigh the intrinsic switching-speed advantage of the larger
devices. This highlights the importance of bitcell geometry at the array
level. The more symmetrical M3D cell aspect ratio can also reduce directional
unbalance in row- and column-wise interconnect loads.

The averaged read/write energy-delay product (EDP) is shown in
Fig.~\ref{fig:edp}. Owing to reduced wire capacitance and improved speed,
the M3D cell lowers EDP by 9.7\% relative to Si-HD. The M3D SRAM incurs
$\sim$35\% higher standby power due to the $V_t$ tuning used to improve
PG drive. Although this tuning increases IGO PG leakage, the increase does not translate proportionally into cell standby power, which depends on the entire 6T network and internal-node/BL biases.
%======================================================================
\subsection{Reliability: BTI-Induced $V_t$ Shift}
\begin{figure*}[p]
  \vspace{-35pt}   % pull the next row up
  \begin{minipage}[t]{0.46\textwidth}\centering
    \includegraphics[width=\linewidth]{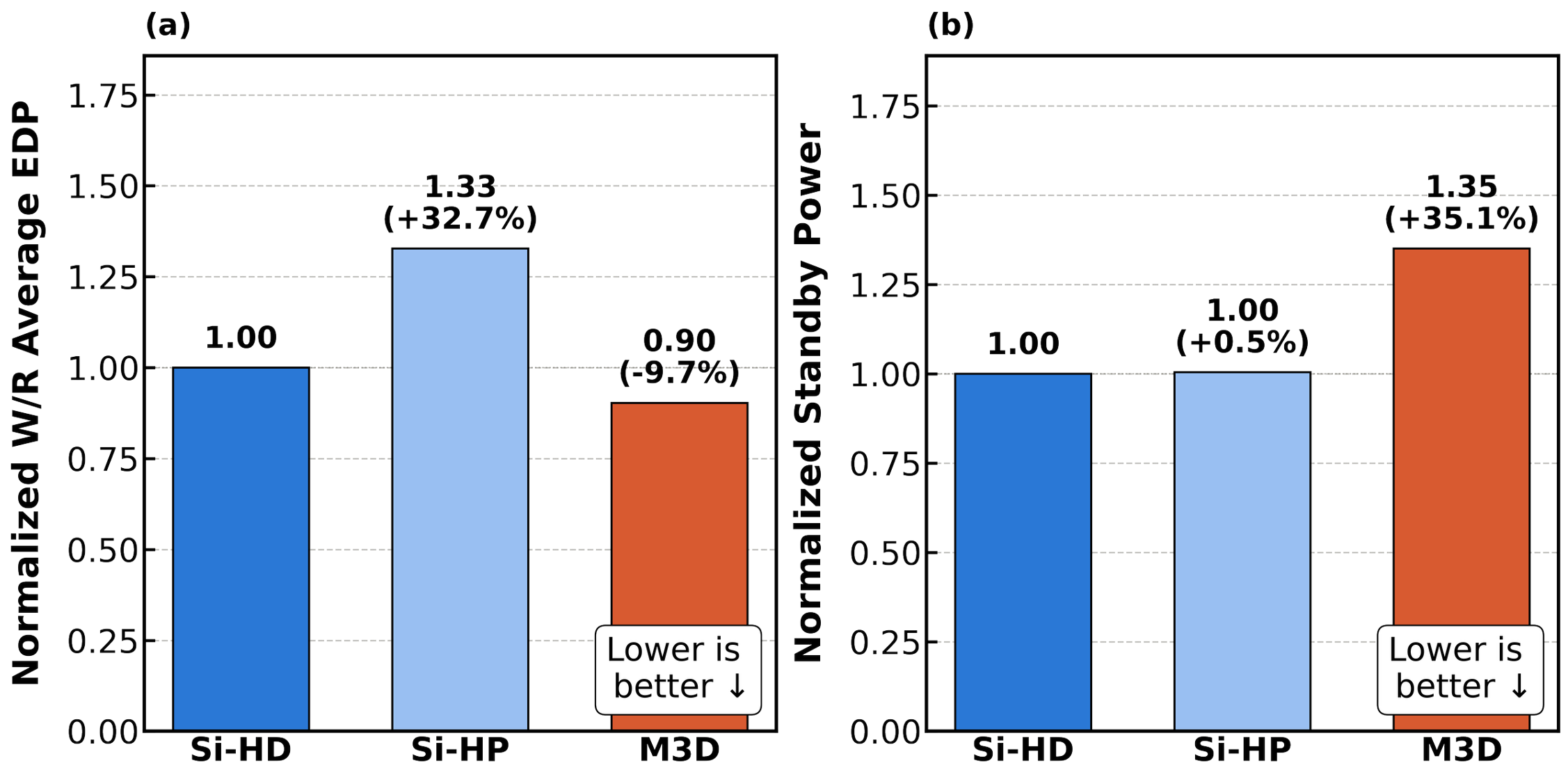}
    \caption{(a) Normalized average W/R EDP and (b) normalized standby power with Si-HD as the baseline. The M3D cell lowers EDP while incurring higher standby power from the $V_t$ tuning used for PG drivability.}
    \label{fig:edp}
  \end{minipage}\hfill
  \begin{minipage}[t]{0.52\textwidth}\centering
    \includegraphics[width=\linewidth]{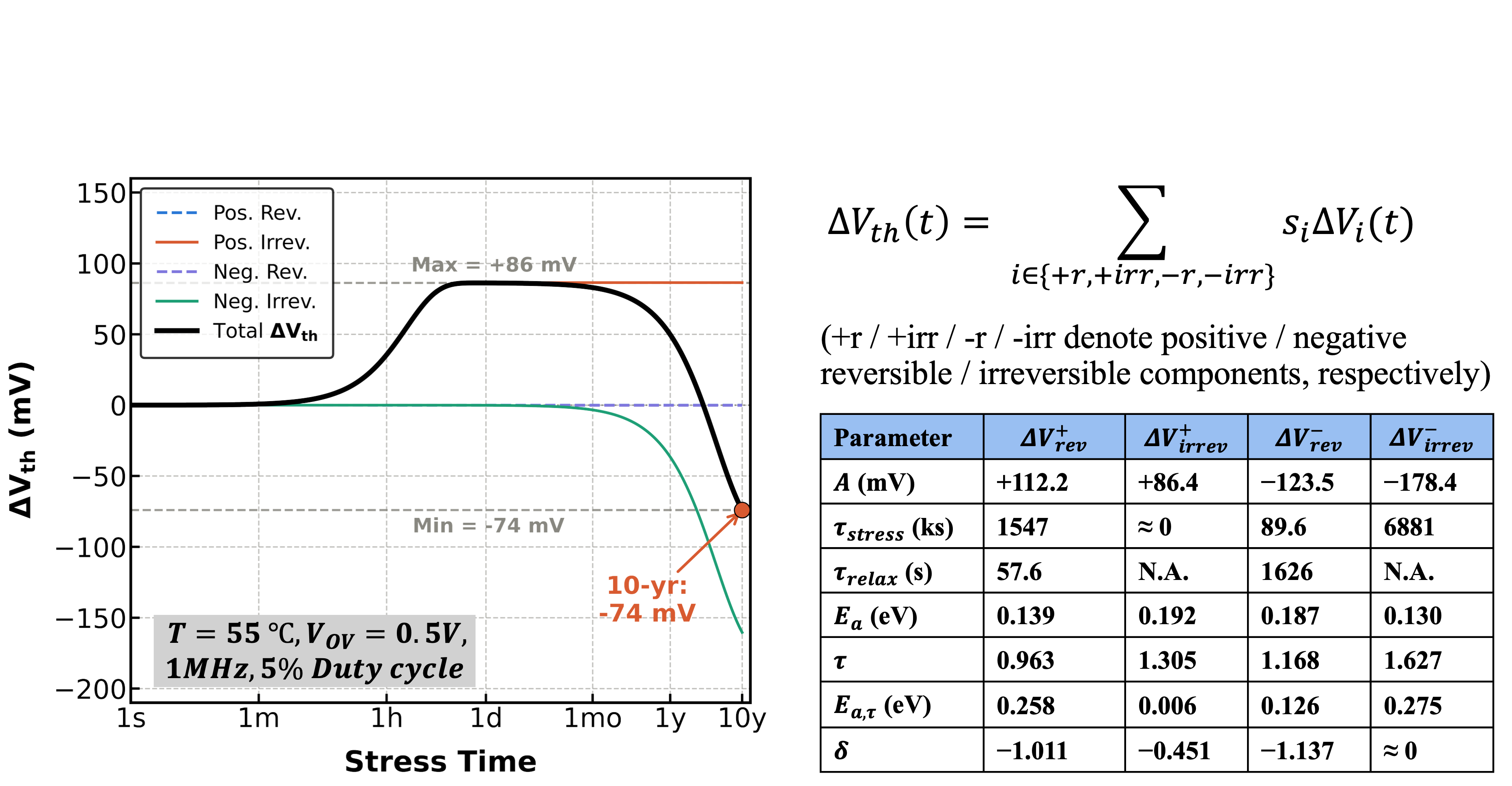}
    \caption{BTI-induced $\Delta V_t$ of the IGO PG vs stress time on a logarithmic scale. The four $\Delta V_{t,i}$ components are summed;
    the total peaks at +86mV and approaches -74mV at 10 years. Table: fitted prefactors and time constants.}
    \label{fig:pbti}
  \end{minipage}\hfill
\end{figure*}

AOSTFTs have long been challenged by bias temperature instability (BTI)
during operation. \cite{sarkar25} demonstrated
suppressed BTI through a tri-layer gate stack and an optimized fabrication
process. To reflect a realistic operating condition, we assume a WL
frequency of 1\,MHz with a 5\% duty cycle at $55^{\circ}$C and an overdrive
of $V_{ov}=0.5$\,V. The measured BTI response is modeled and calibrated using four components, each with its own prefactor and time constant: positive and
negative, reversible and irreversible~\cite{lee25}. The resulting total
$\Delta V_t$ as a function of stress time is shown in
Fig.~\ref{fig:pbti}.

The total $\Delta V_t$ is non-monotonic due to competition among the four BTI components. The two extrema stress the cell in opposite ways:
a positive $\Delta V_t$ weakens the PG and degrades writability,
whereas a negative $\Delta V_t$ increases standby power. We apply both
extrema from Fig.~\ref{fig:pbti} to the subarray testbench and confirm that the
proposed M3D cell remains functional through the projected 10-year lifetime, as shown in Fig.~\ref{fig:bti_vdd}.

%======================================================================
\subsection{Pull-Down Width Trade-off}
The read/write balance depends on the relative strengths of the PU, PG,
and PD devices. With the PU width fixed at 10\,nm, we sweep the PD width,
$W_n$, from 10 to 20\,nm, spanning the conventional 1:1--1:2 PU-to-PD
sizing range of 111 and 122 design points (Fig.~\ref{fig:dtco}).
As $W_n$ increases, RSNM improves while WSNM decreases. The delays show
a similar trade-off: read delay decreases, whereas write delay increases
by approximately 5\%. Thus, the optimum is not determined
by a single metric, but by an application-dependent balance between read/write margins and speed. The proposed M3D cell adopts $W_n=20$\,nm,
which maximizes RSNM while maintaining sufficient WSNM and keeping both read and write delays within 5\% of their respective optimum values.

%======================================================================
\section{\textbf{Conclusion}}\label{sec:concl}
\begin{figure*}[p]
\vspace{-1pt}   % pull the next row up
  \begin{minipage}[t]{0.29\textwidth}\centering
    \includegraphics[width=\linewidth]{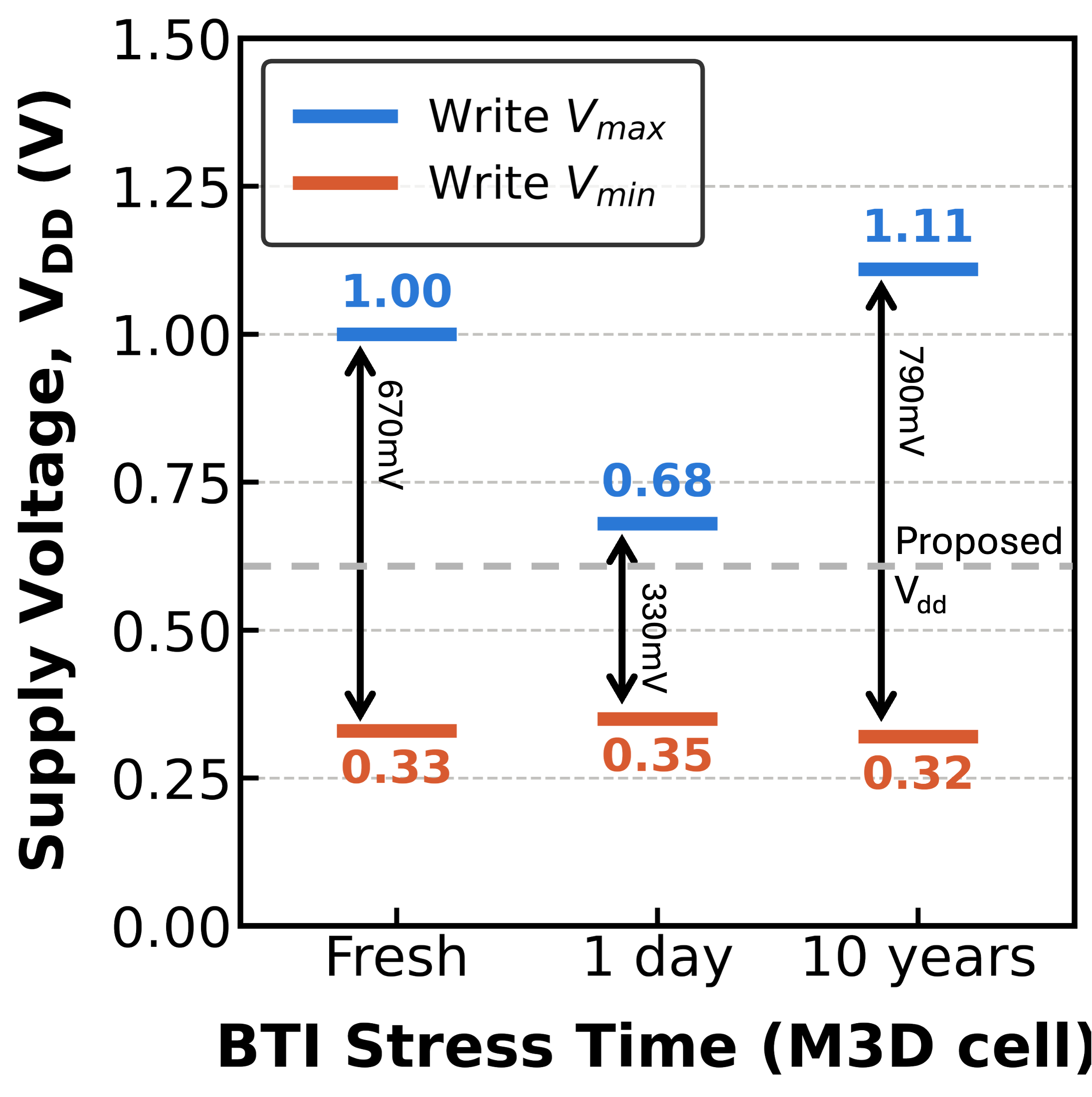}
    \caption{Write-operable $V_{DD}$ window of the M3D cell at selected BTI stress milestones, showing functional operation under the projected PG $V_t$ shifts. The bias-dependent strength ratio between the IGO PG and Si PD makes write operation the limiting condition.}
    \label{fig:bti_vdd}
  \end{minipage}\hfill
  \begin{minipage}[t]{0.34\textwidth}\centering
    \includegraphics[width=\linewidth]{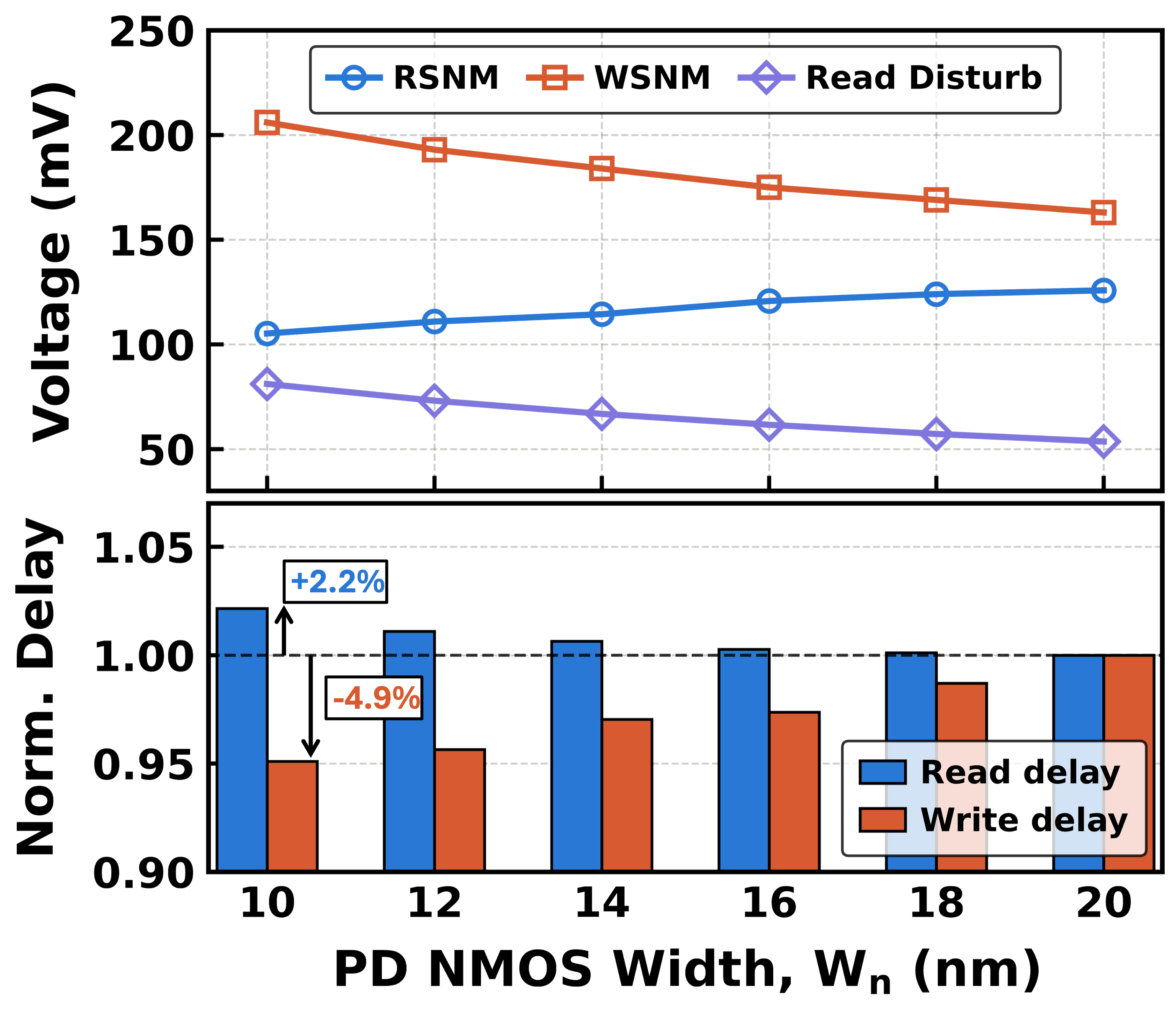}
    \caption{ Co-optimization of the M3D cell as a function of PD NMOS width $W_n$. 
    Increasing $W_n$ strengthens the PD, improving RSNM and read-disturb margin at the cost of WSNM, write delay, and cell area.} 
    \label{fig:dtco}
  \end{minipage}\hfill
  \begin{minipage}[t]{0.31\textwidth}\centering
    \includegraphics[width=\linewidth]{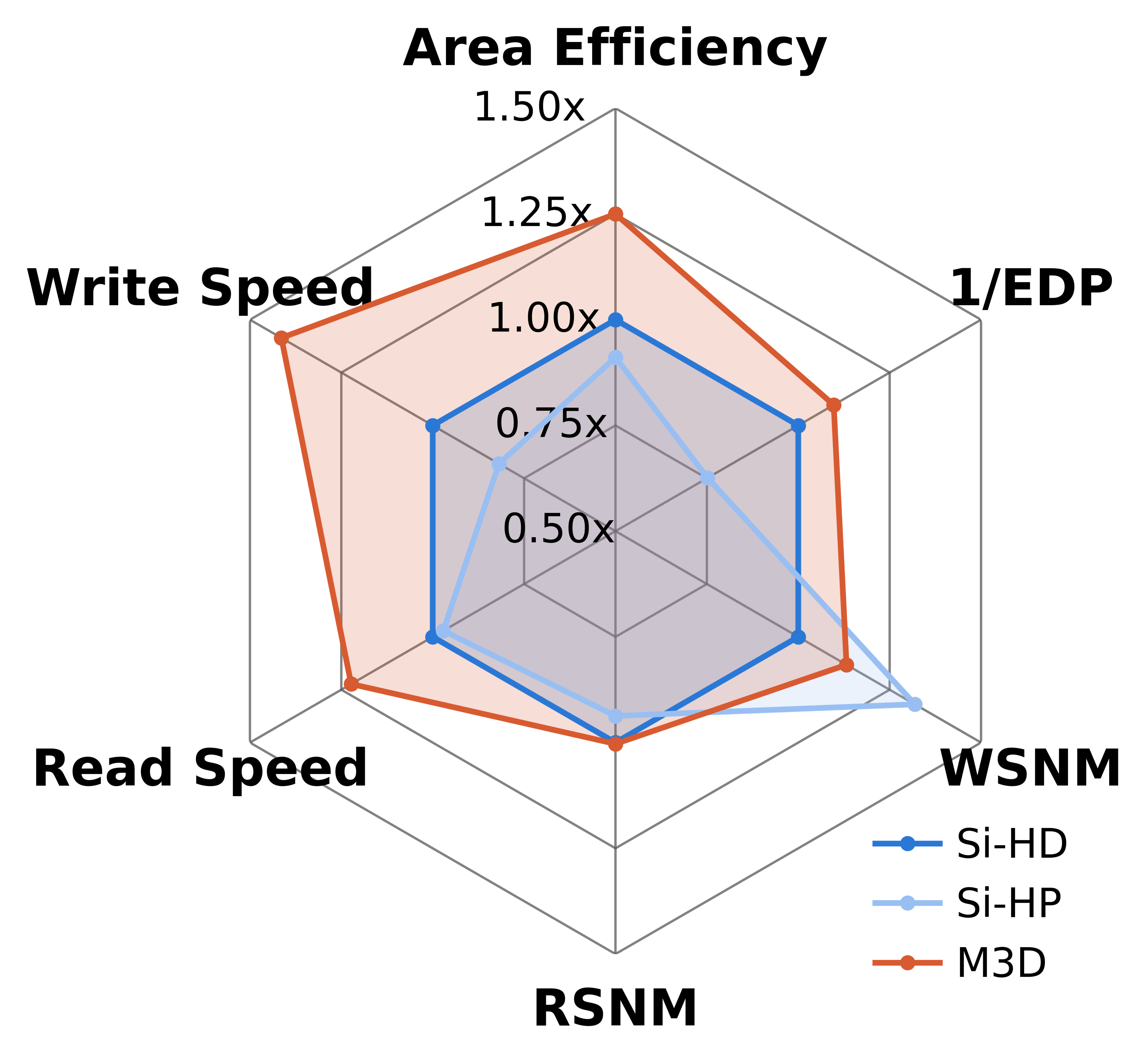}
    \caption{Summary of normalized performance metrics: area efficiency, write speed, read speed, RSNM, WSNM, and EDP. A larger enclosed area of the plot indicates better overall performance.}
    \label{fig:summary}
  \end{minipage}\hfill
\end{figure*}

We present a process-aware design and optimization of a 2nm hybrid M3D
6T SRAM integrating BEOL DG IGO PGs with a Si-nanosheet latch. A
neighbor-cell shared S/D contact mitigates IGO contact resistance and
restores PG drive without area penalty. The optimized cell is 25\%
smaller than the Si-HP baseline while remaining writable over
0.4--1.0\,V. At the 128$\times$256 subarray level, its reduced footprint
and parasitics lower write delay by 42.2\%, read delay by 22.3\%, and
EDP by 9.7\% relative to Si-HD, with a $\sim$35\% standby-power penalty
from $V_t$ tuning. Under projected AC BTI, the PG $V_t$ remains
within $+86$/$-74$\,mV over 10 years, with functional operation at both
extrema. Key performance metrics are summarized in
Fig.~\ref{fig:summary}. These results demonstrate that vertically
integrating the PGs while co-optimizing device, contact, and cell design
provides a promising path toward continued SRAM density scaling beyond
conventional 2D layouts.

\vspace{5pt}
{\small
\noindent\textbf{Acknowledgment:} This work was supported by CoCoSys: Center
for the Co-Design of Cognitive Systems, one of seven centers in JUMP 2.0, an
SRC program sponsored by DARPA.
}

\balance
%======================================================================
% FIGURE BLOCK  --  3 per row, left-to-right then top-down,
% independent Fig. 1-15 numbers, collected on float pages (pp. 4-5).
% Each row is one figure* with three side-by-side minipages; every
% \caption increments the figure counter so numbering stays separate.
%======================================================================

% ---------- ROW 1 : Fig. 1, 2, 3 ----------

% ---------- ROW 3 : Fig. 7, 8, 9 ----------

% ---------- ROW 4 : Fig. 10, 11, 12 ----------

% ---------- ROW 5 : Fig. 13, 14, 15 ----------

%======================================================================

\end{document}